\documentclass[a4paper,fleqn,usenatbib,twocolumn]{mnras}
\setcitestyle{}

\usepackage{newtxtext}
\usepackage[T1]{fontenc}
\usepackage{ae,aecompl}
\usepackage{color}

\usepackage{graphicx}	% Including figure files
\usepackage{amsmath}	% Advanced maths commands
\usepackage{amssymb}	% Extra maths symbols
\title{Very-extreme-mass-ratio bursts in the Galaxy and neighbors for space-borne detectors}% Force line breaks with \\
\author[Wen-Biao Han, Xing-Yu Zhong, Xian Chen, Shuo Xin ]{
Wen-Biao Han$^{1,2,3}$\thanks{E-mail: wbhan@shao.ac.cn}, Xing-Yu Zhong$^{1,2,7}$\thanks{E-mail: zxy@shao.ac.cn}, Xian Chen$^{4,5}$\thanks{E-mail:xian.chen@pku.edu.cn}, Shuo Xin $^{6,8}$\thanks{E-mail: xinshuo@tongji.edu.cn}
\\
 $^1$Shanghai Astronomical Observatory, Shanghai, 200030, China\\
 $^2$School of Astronomy and Space Science, University of Chinese Academy of Sciences, Beijing 100049, China\\
 $^3$School of Fundamental Physics and Mathematical Sciences, Hangzhou Institute for Advanced Study, UCAS, Hangzhou 310024, China \\
 $^4$Astronomy Department, School of Physics, Peking University, Beijing 100871, China\\
 $^5$Kavli Institute for Astronomy and Astrophysics at Peking University, Beijing 100871, China \\
 $^6$Tongji University, Shanghai, 200092, China\\
 $^7$International Centre for Theoretical Physics Asia-Pacific, Beijing/Hangzhou, China \\
 $^8$Key Laboratory for Research in Galaxies and Cosmology, Shanghai Astronomical Observatory, Shanghai 200030, China
 }

\date{Accepted XXX. Received YYY; in original form ZZZ}

\pubyear{2020}

\begin{document}
\label{firstpage}
\pagerange{\pageref{firstpage}--\pageref{lastpage}}
\maketitle

\begin{abstract}
Two recent papers\citep{xmri1, xmri2} revealed that in our Galaxy there are very extreme-mass-ratio inspirals composed by brown dwarfs and the supermassive black hole at the center of the Galaxy. The event rates estimated in these papers are very considerable for future space-borne detectors. In addition, there are plunge events during the formation of inspiraling orbits. In this work, we calculate the gravitational waves from compact objects (brown dwarf, primordial black hole and etc.) plunging into or being scattered by the central supermassive black hole. We find that for space-borne detectors the signal-to-noise ratios of these bursts are quite high. The event rates are estimated as $\sim$ $0.01 {\rm{yr}^{-1}}$ for the Galaxy. If we are lucky, this kind of very extreme-mass-ratio bursts will offer a unique chance to reveal the nearest supermassive black hole and nuclei dynamics. The event rate can be as large as 4 $\sim$ 8 ${\rm yr^{-1}}$ in 10 Mpc, and because the signal is strong enough for observations by space-borne detectors, we have a good chance of being able to probe the nature of neighboring black holes.
\end{abstract}

%\tableofcontents
%\section{§1. Introduction}
\section{introduction}

After one century, the gravitational waves (GWs) predicted by Einstein’s gravitational theory have been detected by the advanced Laser Interferometer Gravitational-wave Observatory (aLIGO) and advanced Virgo (AdV), with more than 10 events in the O1 and O2 runs \citep{gw15a,gw15b,gw17a,gw17b}. The success of GW
detection has motivated plans for space-borne interferometers with arm-lengths of about a million kilometers. The Laser Interferometer Space Antenna (LISA \cite{lisa}), Taiji \citep{hu2017the} and Tian-Qin \citep{luo2016tianqin}, with planned launches in the 2030s, will focus the observation band from 0.1 milli-Hertz to 1 Hz. An extreme-mass-ratio inspiral (EMRI) \citep{second_editor_1,emris}, for example a stellar-mass compact object (1-10 ${\rm M_\odot}$) orbiting around a supermassive black hole (SMBH), is a promising source of a GW signal in the band of these space-borne detectors \citep{second_editor_1,second_editor_2,lisal3,second_editor_3}. In particular, there are extreme mass-ratio bursts (EMRBs), which are produced when a compact object passes through periapsis on a highly eccentric orbit about a much more massive object \citep{rubbo06,hopman07,2008ApJ,toonen09,  berry13a, berry13b,2017MNRAS,2019MNRAS}. The event rate of this kind of burst source is $0.2 {\rm {yr}^{-1}}$ within 100 Mpc, and the signal-to-noise ratio is up to a few tens based on the analysis by \citep{berry13b}.

Recently, two groups independently reported that in our Galactic Centre, LISA will see a few very extreme mass-ratio inspirals. The mass ratio is about $10^{-8}$ \citep{xmri1, xmri2}, and this kind of source, called a X-MRI \citep{xmri1} is composed of a brown dwarf inspiralling into the SMBH. The event rates these authors estimated are quite high, and could be more than 10 once the LISA starts to observe.

There are, however,many more brown dwarfs with unbound orbits than with bound ones (X-MRIs). These brown dwarfs with plunge orbits will collide with the SMBH and produce burst GWs, if the event rate is considerable, LISA will detect this kind of GWs. In the
present paper, we refer to this kind of GW source as a very extreme mass-ratio burst (XMRB). In our Galaxy, the signal-to-noise ratio (SNR) of XMRBs will be as large as a few thousand. XMRBs in the Galactic Centre will be very easy to find with LISA, and the GWs of
XMRBs will give us a unique chance to investigate the nature of the central SMBH. In this paper, we first calculate the waveforms and SNRs of some typical XMRBs, and then we show that the event rate of this kind of GW burst is about $10 {\rm {yr}^{-2}}$ in our Galaxy and about $4-8 {\rm {yr}^{-1}}$ within 10 Mpc distance. Finally, we address the expectation
of the detection of XMRBs in our Galaxy and neighboring galaxies by LISA, Taiji and Tian-Qin.

%\section{§2. Orbits and Waveforms of XMRBs}
%\subsection{2.1. Orbits of XMRBs}
%add Teukolsky equation here shortly.
\section{Orbits and Waveforms of XMRBs}

An XMRB is composed of a compact object such as a brown dwarf or primordial black hole with mass $10^{-2} \sim 10^{-1} {\rm M_\odot}$, and in our Galaxy the central black hole has a mass of about $4.3\times 10^{6} {\rm M_\odot}$. The small objects can thus be safely treated as test particles. For an unbound orbit of a test particle plunging into or scattered by a Schwarzschild black hole, the geodesics are governed by the following equations (with geometric units: $c=G=1$):

%
%and we can get the dynamic euaqtion:
%
\begin{align}
(\frac{dr}{dt})^2&=[E^2-(1-\frac{2M}{r})(1+\frac{L^2}{r^2})]\cdot {(1-\frac{2M}{r})}^{2}\cdot E^{-2}\\
\frac{d\varphi}{dt}&=\frac{L}{r^2}\cdot(1-\frac{2M}{r})\cdot E^{-1}
\end{align}

\begin{figure}
\centering
\includegraphics[width=\linewidth,scale=1.00]{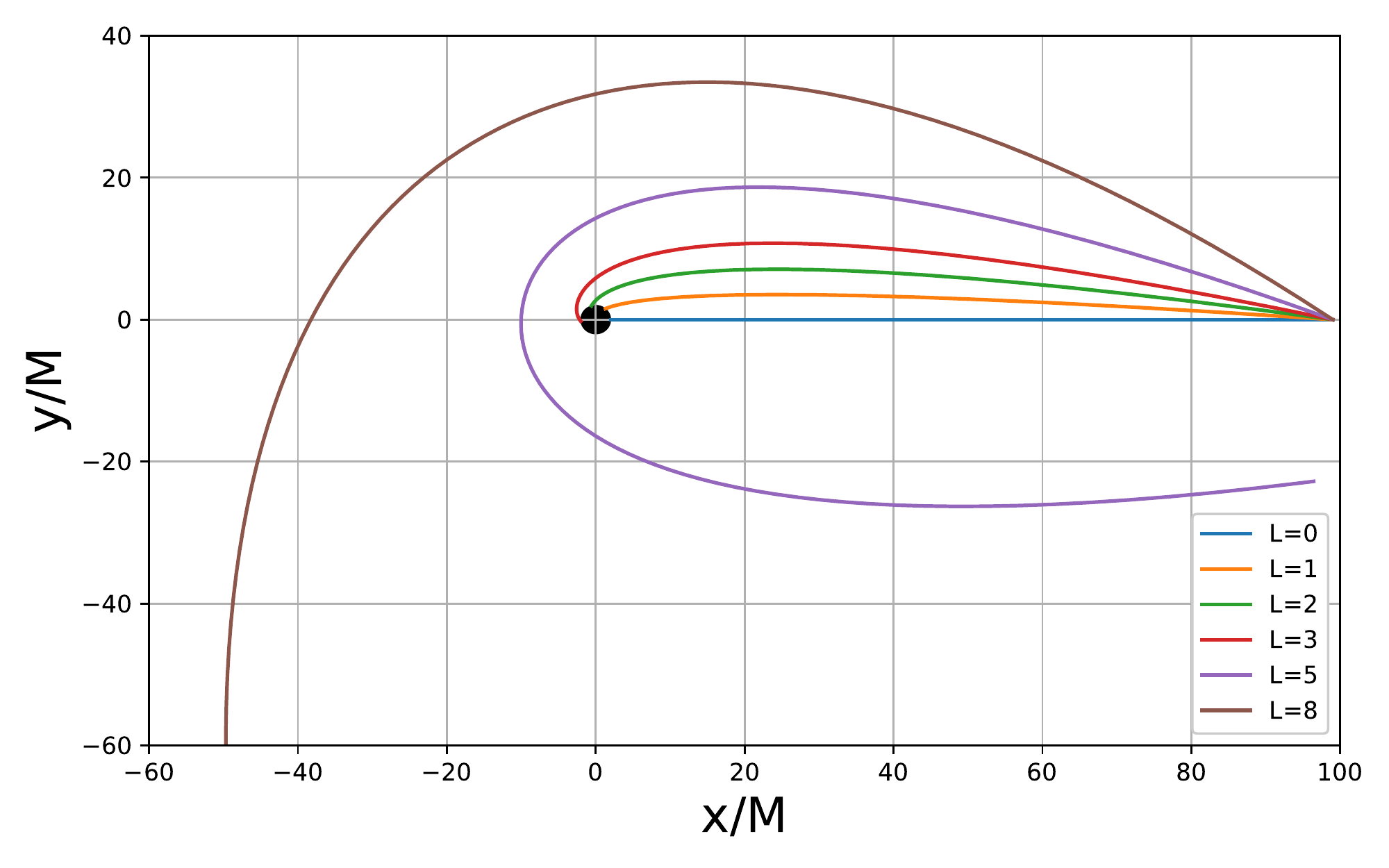}
\caption{The orbits of compact objects with different angular momentum starting from 99 $M$ to the central supermassive black hole.}
\label{orbit}
\end{figure}

The initial velocity is assumed to be much lower than the velocity of light, then the energy of particle $E=1$ in the unit of the mass itself, the valid potential energy $U^2=(1-\frac{2M}{r})(1+\frac{L^2}{r^2})$ will have a peak value. This value is equal to 1 when $L=4$. So if $0\leqslant L\leqslant 4$ the particle will plunge into the horizon, and if $L>4$, the particle will be scattered. Figure \ref{orbit} shows the plunging and scattered orbits with $L<4$ and $L>4$ respectively.

For the plunging case, we use a formalism based on the Teukolsky equation \citep{Teukolsky} to compute the GW. The Teukolsky equation is a curvature perturbation method for a Kerr black hole and coincides with Regge-Wheeler and
Zerilli equation \citep{RW1957,Z1970} when the spin $a = 0$. Here, we mainly calculate the Schwarzschild cases, but in the future we will use the Teukolsky equation to simulate spinning SMBHs in detail. The numerical method in the frequency domain has been developed in a number of studies \citep{SN,Teukolsky 1996,Hughes 2000, MST_r,MST_c,han10, han17} (and references inside). The perturbation field $\psi_{4}$, decomposed in frequency domain, is

\begin{align}
\psi_{4}=\frac{1}{(r-iacos\theta)^4}\int^{\infty}_{-\infty}d\omega\sum\limits_{lm}R_{lm\omega}(r)_{-2}S^{a\omega}_{lm}(\theta)e^{-i\omega t+im\phi}
\end{align}
where The function $_{-2}S^{a\omega}_{lm}(\theta)$ is a spin-weighted spheroidal harmonic that can be computed via eigenvalue \citep{Hughes 2000} or continuous fraction methods \citep{Leaver 1985}. The radial function $R_{lm\omega}(r)$ obeys the Teukolsky equation:
\begin{align}
\Delta^{2}\frac{d}{dr}(\frac{1}{\Delta}\frac{R_{lm\omega}}{dr})-V(r)R_{lm\omega}(r)=-\mathcal{T}_{lm\omega}(r) \,,
\end{align}
where $\mathcal{T}_{lm\omega}(r)$ is the source term, and $\Delta =r^2-2Mr+a^2$. The potential is $
V(r)=-\frac{K^2+4i(r-M)K}{\Delta}+8i\omega r+\lambda$,
where  $K=(r^2+a^2)\omega-ma$, $\lambda\equiv\varepsilon_{lm}-2am\omega+a^2w^2-2$, where the number $\varepsilon_{lm}$ is the eigenvalue of the spheroidal harmonic.

Using the Green function method \citep{Green}, we can obtain the solution of the Teukolsky equation with a purely outgoing property at infinity and a purely ingoing property at the horizon:
\begin{align}
R_{lm\omega}(r)=&\frac{1}{2i\omega C^{trans}_{lm\omega}B^{inc}_{lm\omega}}\{R^{\infty}_{lm\omega}(r)\int^{r}_{r_{+}}dr'R^{H}_{lm\omega}T_{lm\omega}\Delta^{-2}
\notag
\\
&+R^{H}_{lm\omega}(r)\int^{\infty}_{r_{+}}dr'R^{\infty}_{lm\omega}T_{lm\omega}\Delta^{-2}\}
\end{align}

The asymptotic behavior of this solution near horizon and infinity is
\begin{align}
R_{lm\omega}(r\to r_{+})&=\frac{B^{trans}_{lm\omega}\Delta^{2}e^{-iPr^{*}}}{2i\omega C^{trans}_{lm\omega} B^{inc}_{lm\omega}}\int^{\infty}_{r_{+}}dr'R^{\infty}_{lm\omega}T_{lm\omega}\Delta^{-2}\notag
\\&\equiv Z^{\infty}_{lm\omega}\Delta^{2}e^{-iP r^{*}}
\\R_{lm\omega}(r\to \infty)&=\frac{r^{3}e^{i\omega r^{*}}}{2i\omega B^{inc}_{lm\omega}}\int^{\infty}_{r_{+}}dr'R^{H}_{lm\omega}T_{lm\omega}\Delta^{-2}\notag
\\
&\equiv Z^{H}_{lm\omega}r^{3}e^{i\omega r^{*}}
\end{align}
where $P=\omega-ma/2Mr_{+}$, and $r^{*}$ is the tortoise coordinate.

In general, because the homogeneous solution will diverge near the infinity, we cannot derive solutions directly from the Teukolsky equation with any kind of accuracy. To solve this problem, we can convert the equation to the Sasaki–Nakamura equation \citep{SN}:
\begin{align}
\frac{d^2X_{lm\omega}}{d{r^{*}}^2}-F(r)\frac{dX_{lm\omega}}{d{r^{*}}}-U(r)X_{lm\omega}=0
\end{align}
and use the transform rlue \citep{t-s2,t-s1} from the Sasaki-Nakamura function to the Teukolsky function:
\begin{align}
R^{H,\infty}_{lm\omega}=\frac{1}{\eta}[(\alpha+\frac{\beta_{,r}}{\Delta})\chi^{H,\infty}_{lm\omega}-\frac{\beta}{\Delta}\chi^{H,\infty}_{lm\omega,r}]
\end{align}
where$\chi^{H,\infty}_{lm\omega}=X^{H,\infty}_{lm\omega}\Delta/\sqrt{r^2+a^2}$; $\alpha$, $\beta$, $\eta$ and
the potentials $F(r)$, $U(r)$ can be found in \citep{t-s1}.
In this way,  we can calculate the solutions of the homogeneous Teukolsky equation.

$\psi_{4}$ is related to the amplitude of the GW at infinity as
\begin{align}
\psi_{4}(r\to\infty)\to \frac{1}{2}(\ddot{h}_{+}-i\ddot{h}_{\times}) \,.
\end{align}
The gravitational waveform, observed from distance $R$, latitude angle $\theta$ and azimuthal angle $\phi$, is then given by
\begin{align}
h_{+}-i h_{\times}=\frac{2}{R}\sum\limits_{lm}\int^{\infty}_{-\infty}d\omega \frac{1}{\omega^{2}}Z^{H}
_{lm\omega-2}S^{a\omega}_{l m}(\theta)e^{i(m\phi-\omega[t-r^{*}])}
\end{align}
%
%
%\subsection{2.3. Waveforms of  XMRBs}
Now, we compute the waveforms for the XMRBs with angular momentums $L=0$, $L=1$, $L=2$, $L=3$. Because $L<4$, all these XMRBs will plunge into the black hole directly. These waveforms are calculated using the frequency-domain Teukolsky equation detailed
above.

\begin{figure}
\centering
\includegraphics[width=0.49\linewidth,scale=1]{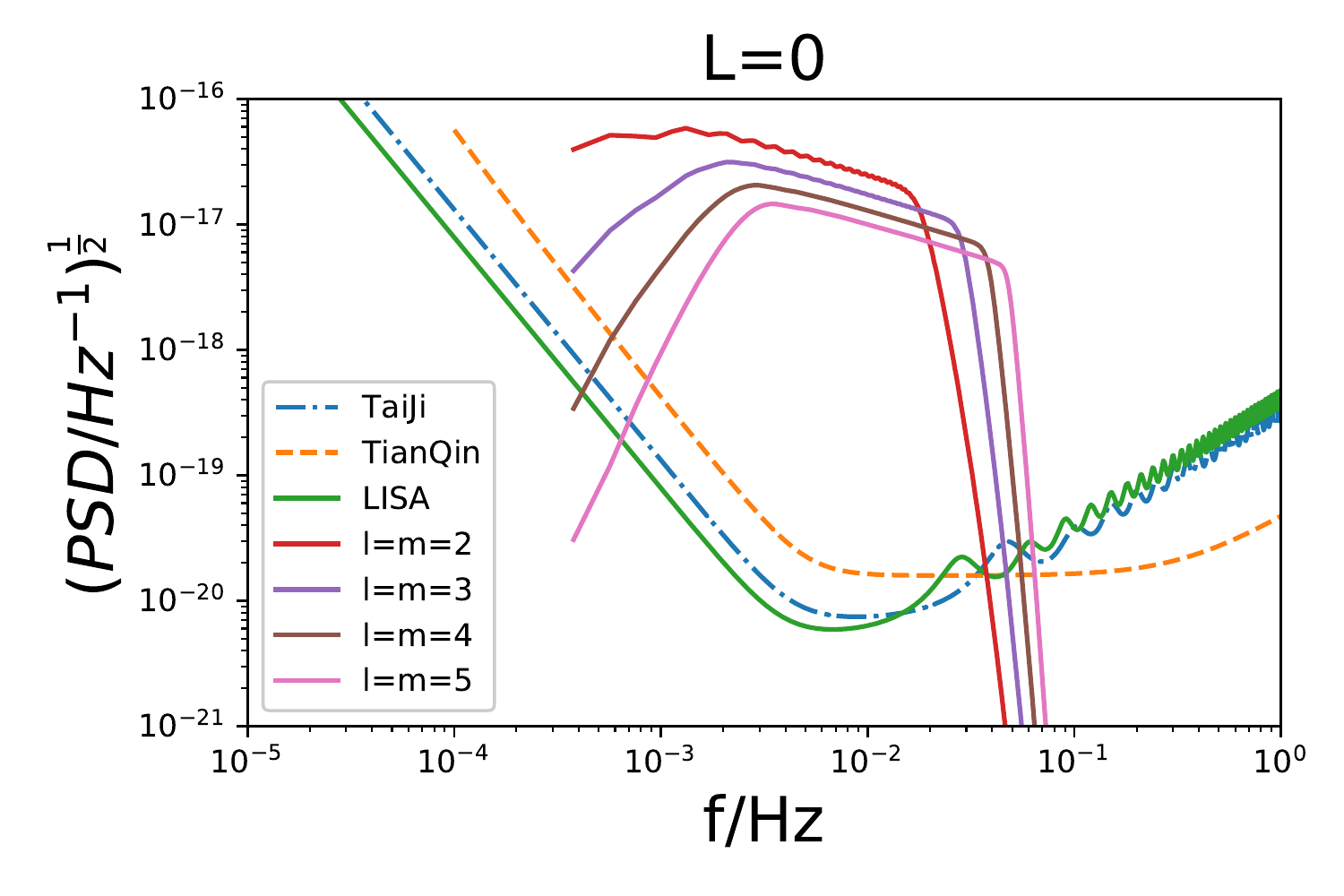}
\includegraphics[width=0.49\linewidth,scale=1]{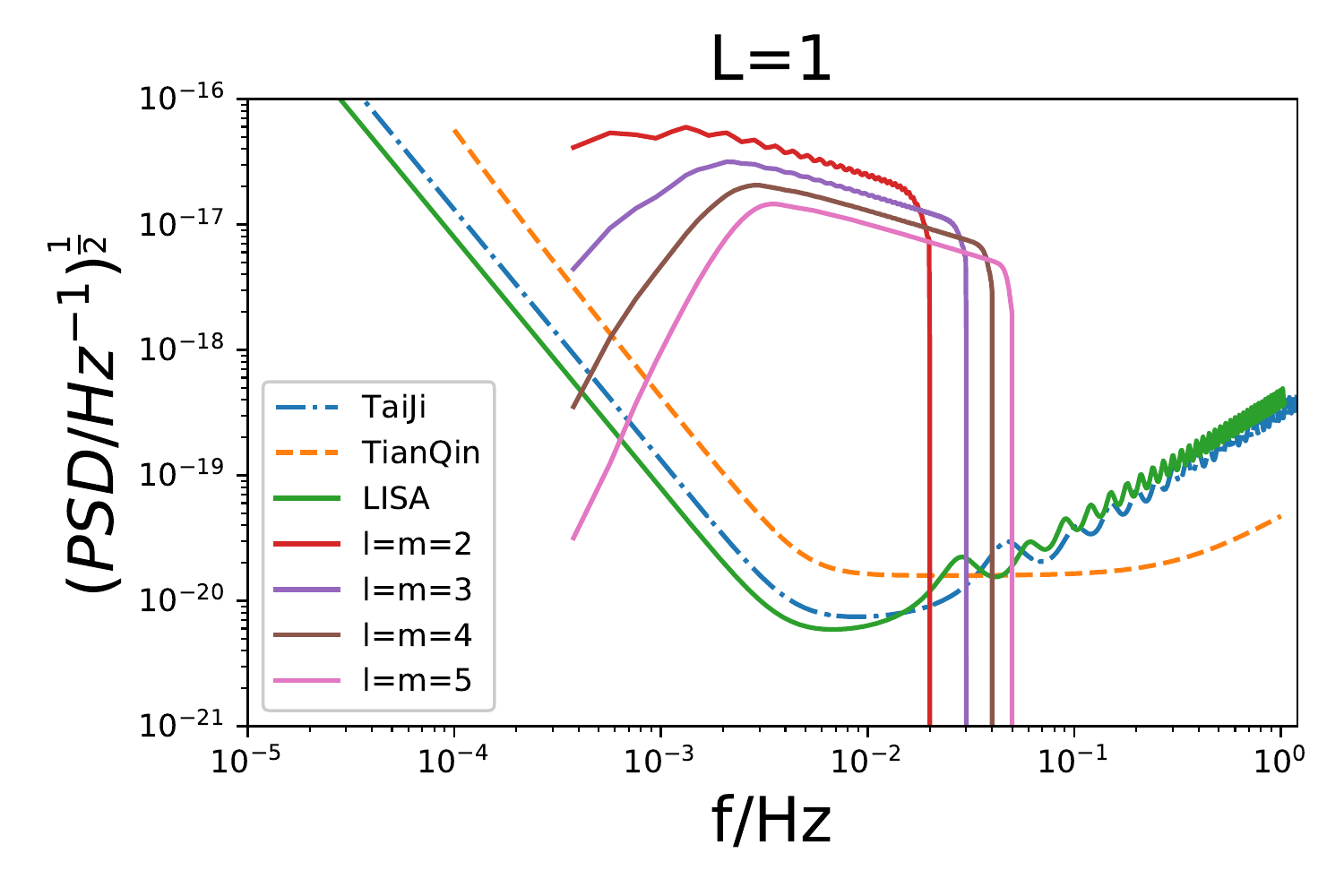}
\includegraphics[width=0.49\linewidth,scale=1]{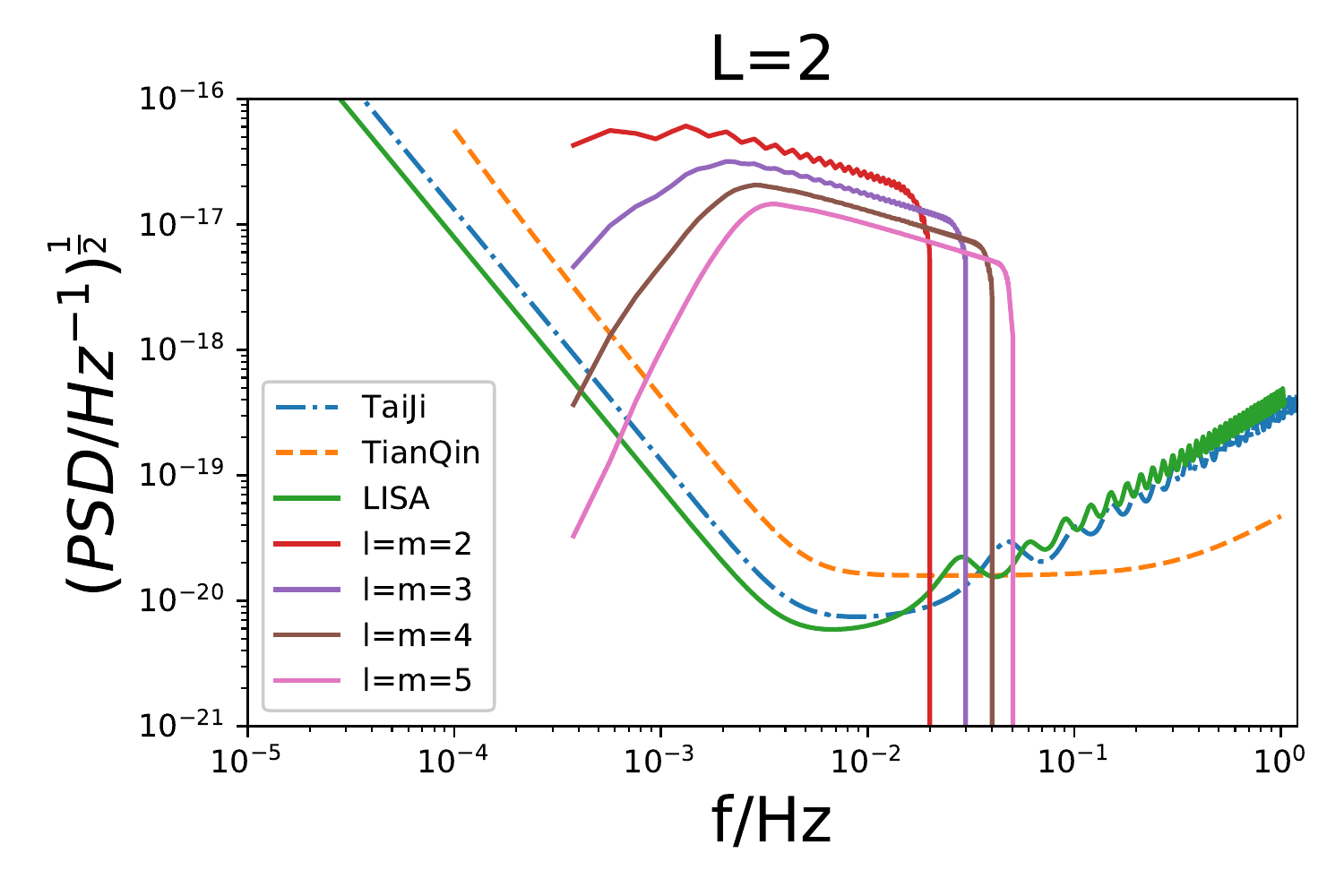}
\includegraphics[width=0.49\linewidth,scale=1]{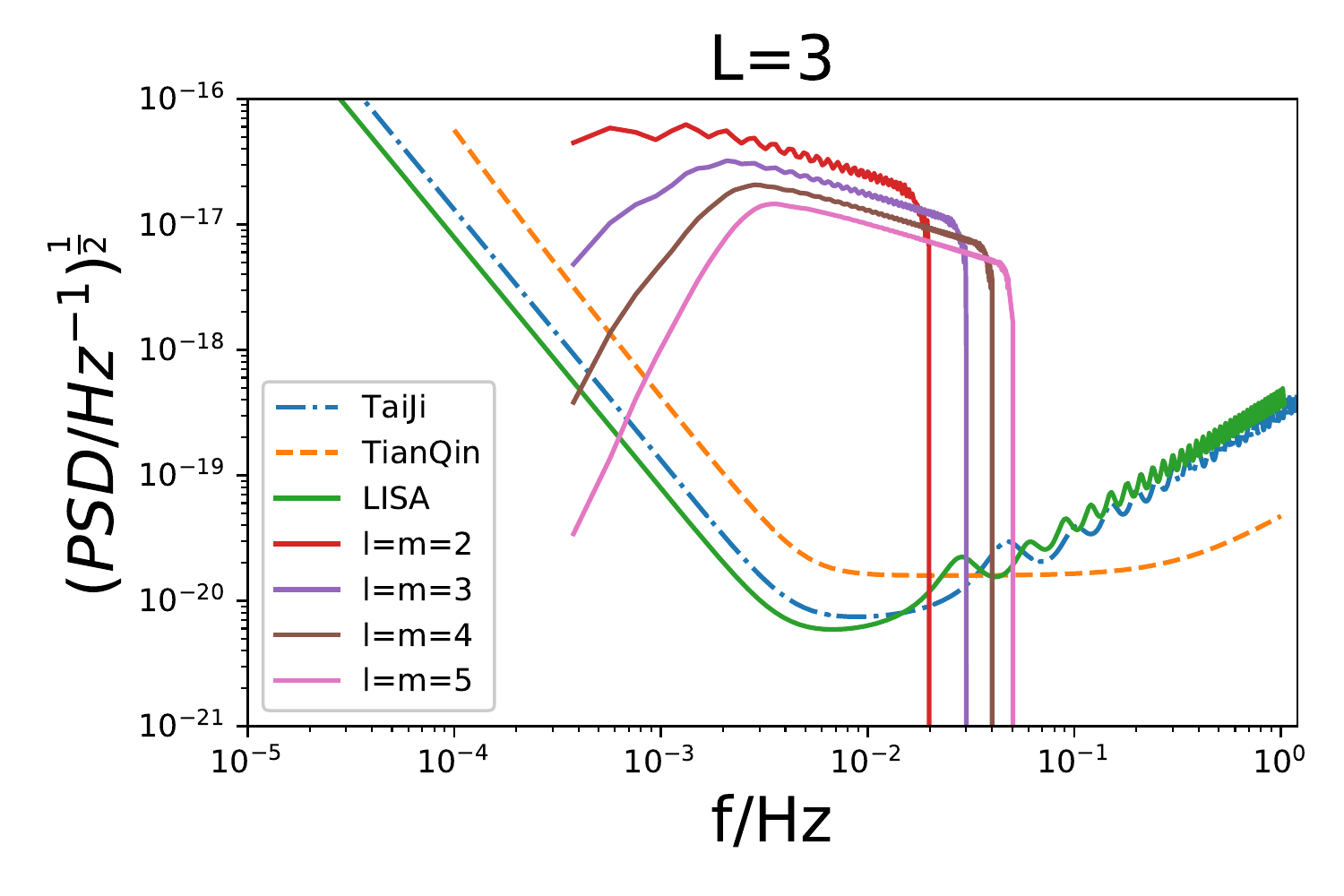}
\caption{Sensitivity curves (LISA, Taiji and Tianqin) and XRMB's GW amplitude spectral density with the different angular momentum $L=0$, $L=1$, $L=2$, $L=3$. The mass of the plunging object is $0.1 {\rm M_\odot}$.}
\label{f-waveform}
\end{figure}

Figure \ref{f-waveform} shows the very strong GW burst signals produced by the XMRBs in the Galactic Centre, the frequency is about $10^{-2}$ Hz, corresponding to the most sensitive frequency band of the space-borne detectors. We can see not only the (2,2) modes but also the (3,3) (4,4) and (5,5) ones, as they are strong enough in relation to the sensitivity curves of LISA, Taiji, and Tianqin.

\begin{figure}
\centering
\includegraphics[width=0.49\linewidth,scale=1]{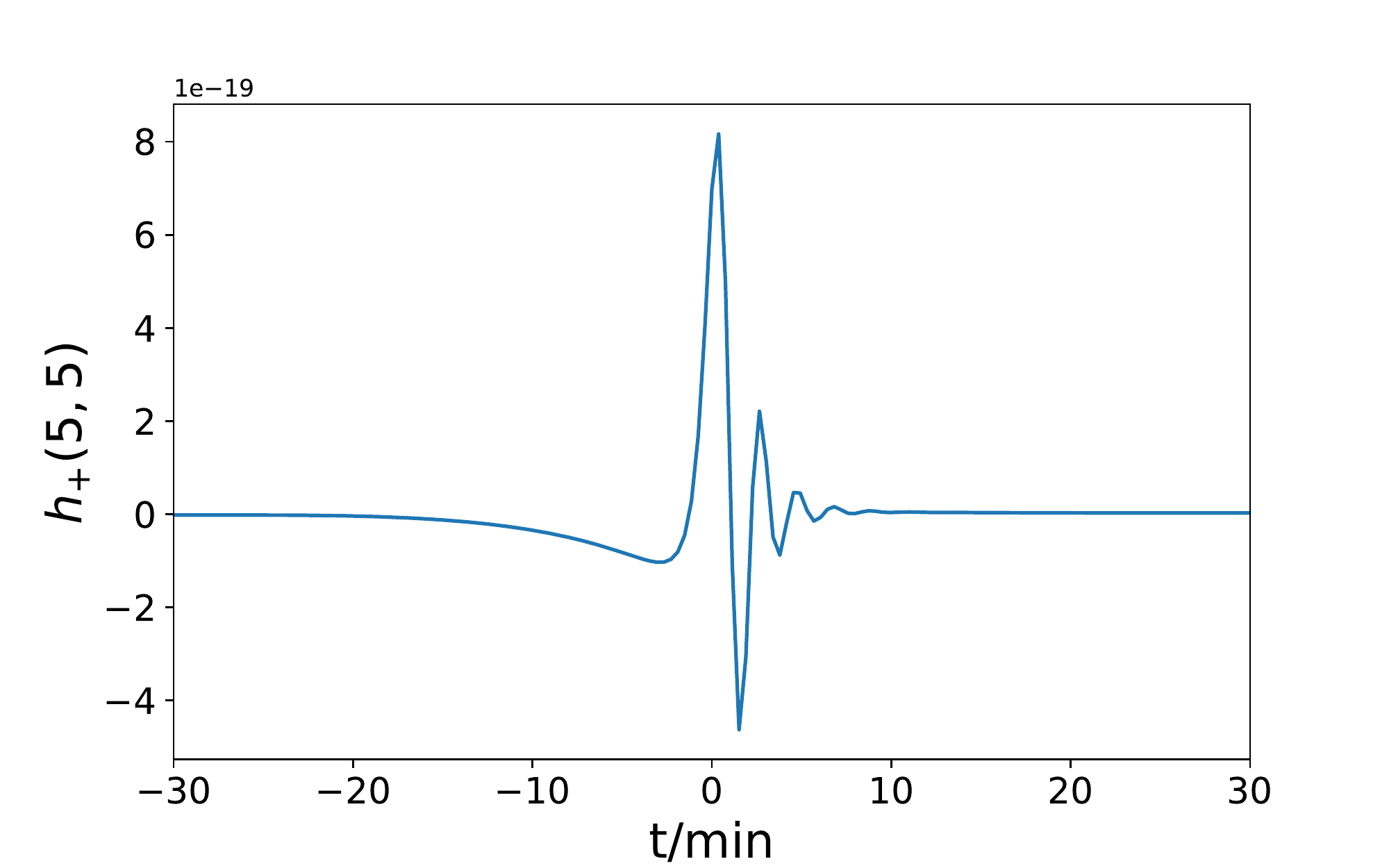}
\includegraphics[width=0.49\linewidth,scale=1]{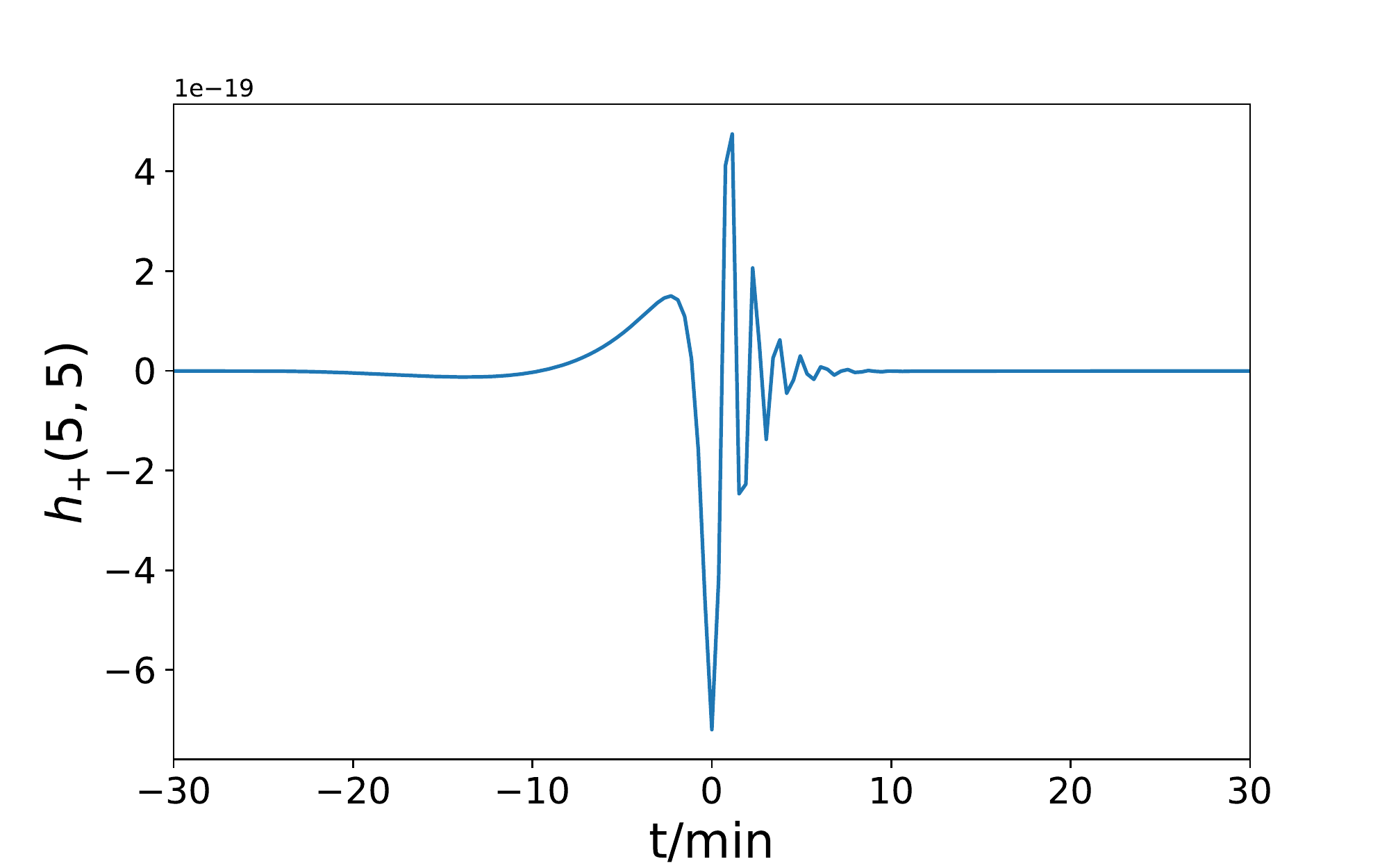}
\includegraphics[width=0.49\linewidth,scale=1]{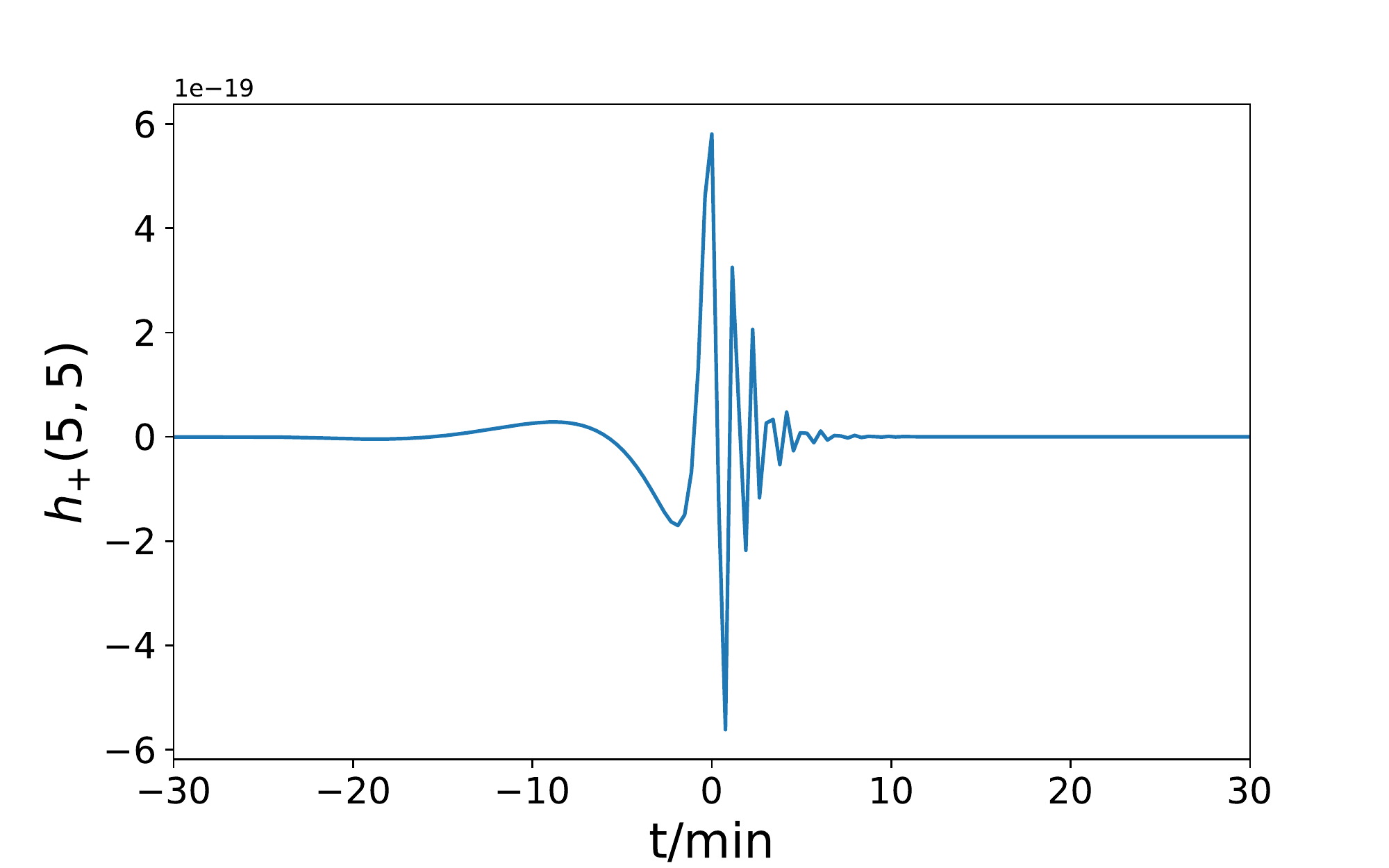}
\includegraphics[width=0.49\linewidth,scale=1]{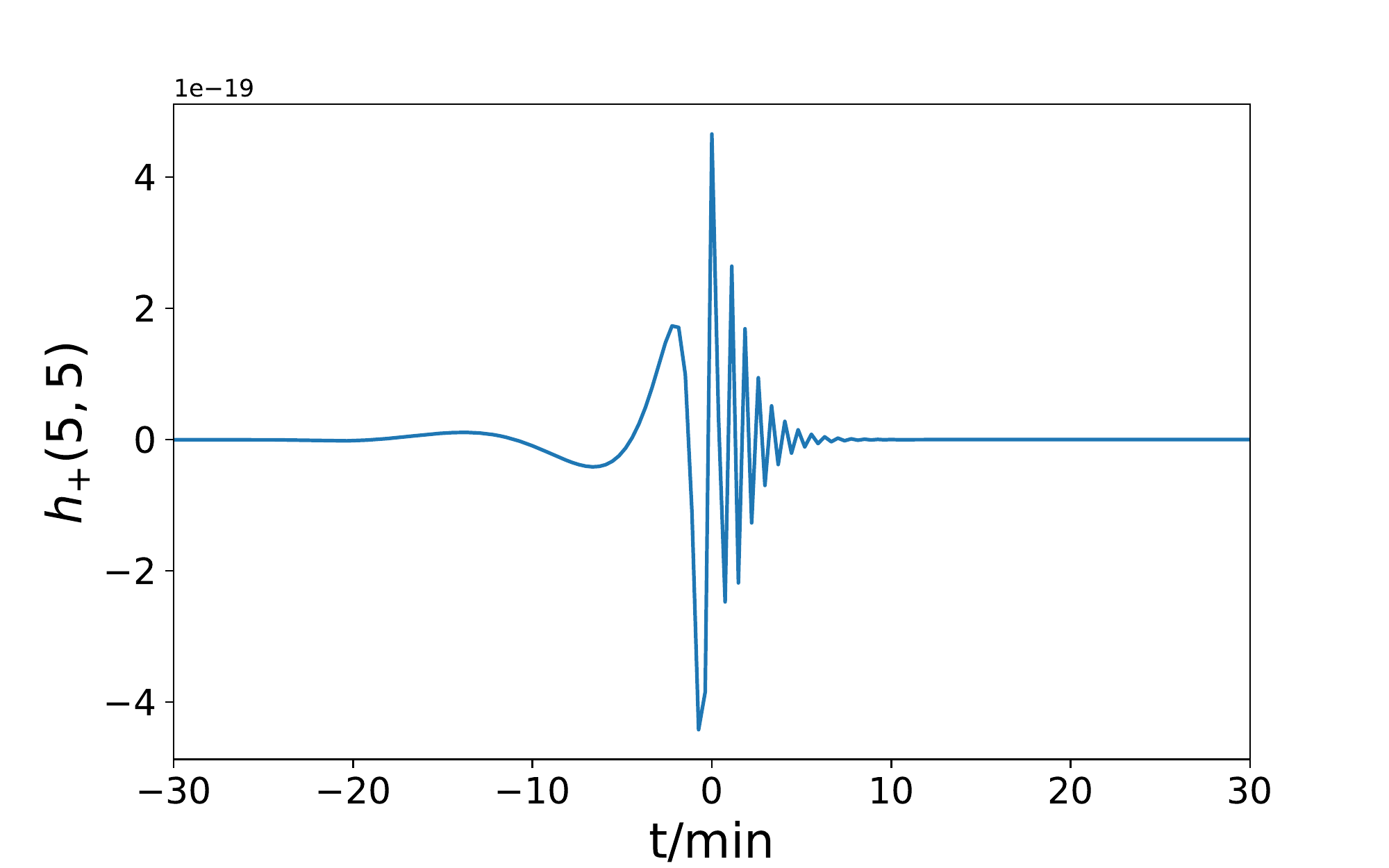}
\caption{The time-domain waveforms of the different harmonic modes for the $L=0$ case.}
\label{t-waveform_L=0}
\end{figure}

Figure \ref{t-waveform_L=0} shows the time-domain waveform $h_{+}$ obtained by the inverse Fourier transform: the duration of these signals is around 20 min for the modes (2,2), (3,3), (4,4) and (5,5).

The plunging signals are very strong, and very important for detecting the structure of the central black hole. However, there are also scattered orbits when the angular momentum $L$ is larger than 4. Because the particle is far away from the central SMBH for its entire orbit (The closest distance is more than 30 $M$ from the SMBH for the $L = 8$ orbit), and integrating the Teukolsky equation is CPU-expensive, we use the quadrupole formula to obtain the time-domain waveform \citep{peters 1963,peters 1964}, and then obtain the frequency-domain waveform by a Fourier transform. For the $L = 5$ orbit, when the particle passes by the SMBH, the distance can be as close as 10 $M$, and thus the quadrupole formula may be inaccurate, but it is still good enough to give a qualitative result, which is the aim of this paper. However, the scattered orbits are less important than the plunging orbits: the latter ones have a much higher SNR and produce the ringdown signal that is a crucial signature of for a black hole. For completeness, we include the scattered sources (which have almost no influence on the event rate – see the next section) and use the quadrupole approximation to show the signals qualitatively. In addition, the semirelativistic approach developed by \citep{1981Ruffini} also works on the burst waveforms, and should be suitable for parameter estimation.
\begin{figure}
\centering
\includegraphics[width=0.49\linewidth,scale=1]{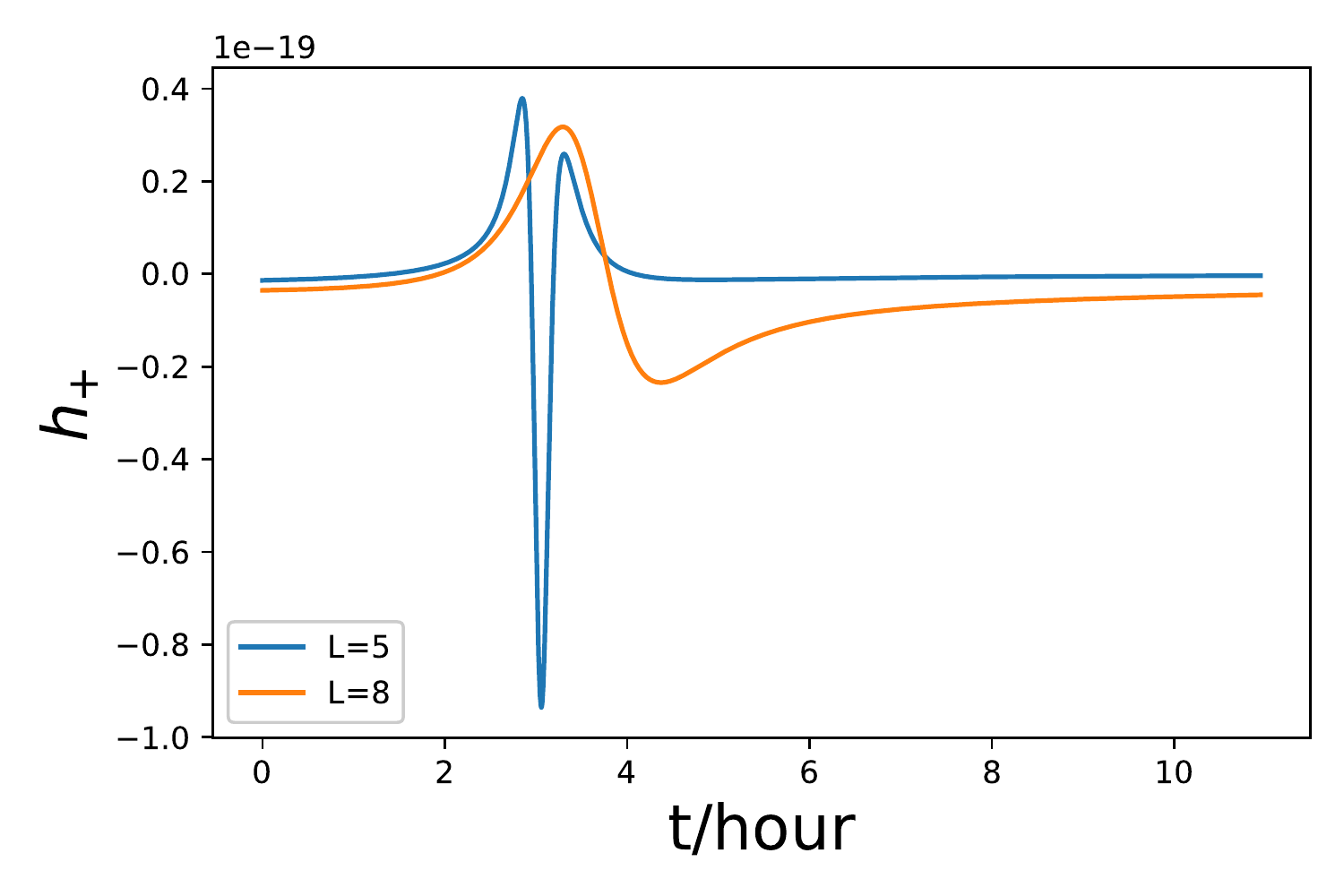}
\includegraphics[width=0.49\linewidth,scale=1]{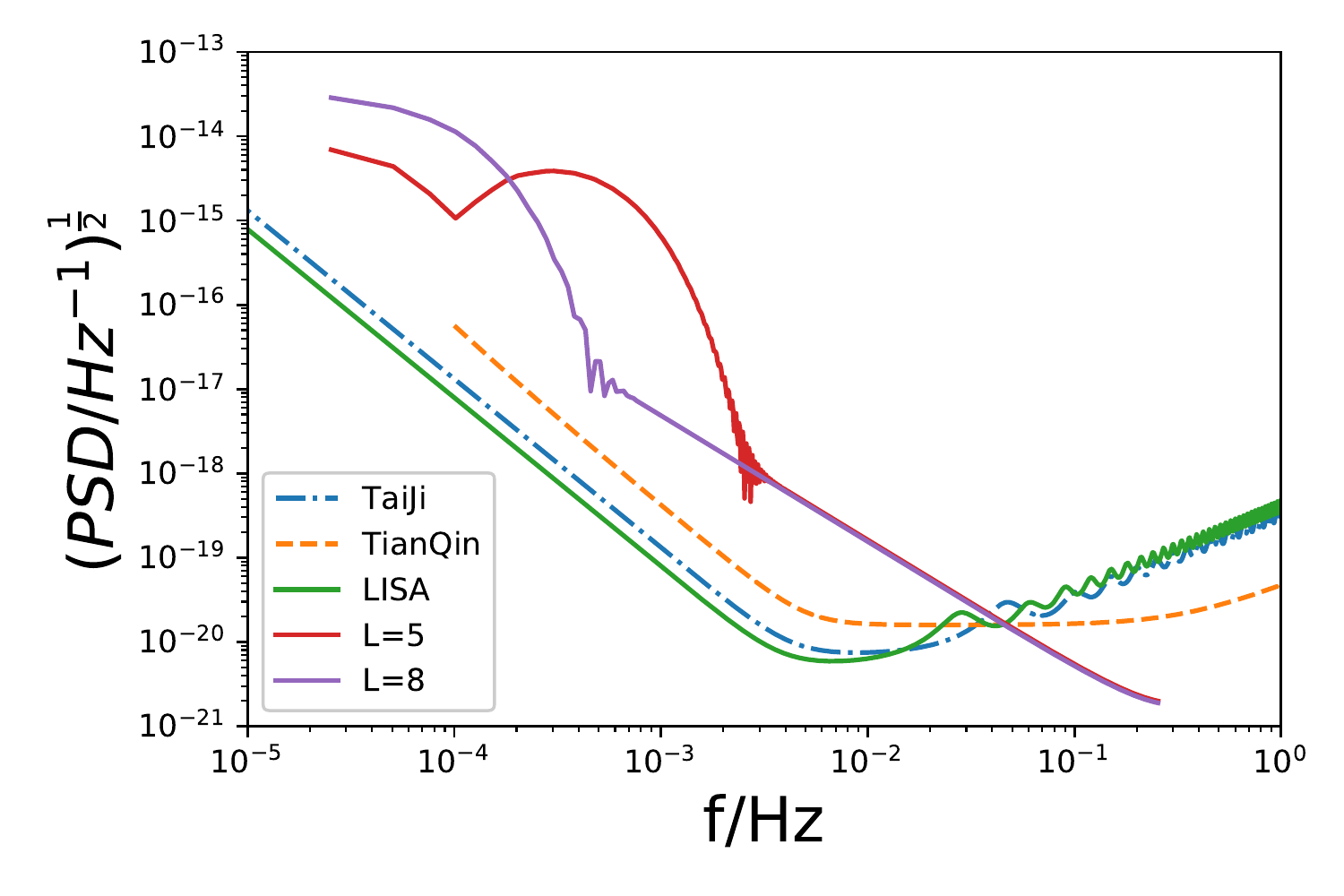}
\caption{time-domain and frequency-domain waveforms for scattered orbits with angular momentum $L=5$, $L=8$, the mass of the plunging object is still $0.1 {\rm M_\odot}$.}
\label{58}
\end{figure}

Figure \ref{58} shows that the signal produced by a scattered particle will exist for several hours. Of course, it is much weaker than the plunging one, but in our Galaxy this kind of signal can still be detected.
%\subsection{2.4. SNRs of GW Signals}

In order to quantitatively demonstrate the strength of the XMRBs,
we compute the SNRs of these signals \citep{SNR}:
\begin{align}
{\rm SNR}^2=4{\rm Re}\int_{0}^{\infty}\frac{|\tilde{h}(f)|^2}{S_n(f)}df
\end{align}
\begin{table*}
\centering
\caption{SNRs of the plunging and scattering GW burst signals in the Galactic center for the space-borne detectors (LISA,TaiJi and Tianqin), the mass of small object is 0.1 solar mass.}
\tabcolsep 0.05in
\begin{tabular}{|c|c|c|c|c|c|c|c|c|c|c|c|c|c|c|c|c|c|c|}
\hline\hline
$L$&\multicolumn{4}{c|}{0}&\multicolumn{4}{c|}{1}&\multicolumn{4}{c|}{2}&\multicolumn{4}{c|}{3} &5&8\\
\hline
$l=m$ &2&3&4&5&2&3&4&5&2&3&4&5&2&3&4&5&2&2\\ \hline
LISA &6468&4430&3193&2392&6477&4435&3197&2392&6495&4444&3202&2397&6527&4457&3211&2401&280&20\\ \hline
TaiJi &4784&3381&2472&1876 &4793&3384&2475&1879&4807&3390&2479&1882 &4830&3400&2484&1886&167&12\\ \hline
TianQin &1965&1445&1091&851&1970&1450&1091&851&1970&1450&1095&851&1984&1454&1095&851&51&4\\
\hline\hline \label{tablesnr}
\end{tabular}
\end{table*}
From Table \ref{tablesnr},  it can be seen that the SNRs of the plunging sources are very high. For the scattered sources, however, the SNR is much lower. For the $L=5$ and $L=8$, the SNRs are 280, 167, 51 and 20, 12, 4 for LISA, Taiji and Tianqin, if in our Galaxy. Therefore, the signals of these scattered sources in neighboring galaxies will be too weak to detect. However, for the plunging burst signals, even at 10 Mpc distance, the signal of the (2, 2) mode may still be detected by LISA if the plunging objects have a slightly greater mass.

Because of the high SNR, we expect that the plunging sources will determine the parameters of the central black hole to a high accuracy. To demonstrate this, we use the Fisher matrix \citep{Cutler1994} to estimate the parameters of an XMRB with a large SNR. The errors for parameter estimation can be approximated by the square root of the diagonal elements of the inverse of the Fisher matrix $\Gamma_{ij}$,
\begin{align}
	\Gamma_{ij}=(\frac{\partial h}{\partial \lambda_i}|\frac{\partial h}{\partial \lambda_j})
\end{align}
where $h$ is the GW signal. In this paper, we consider only two parameters $\lambda_i=(M,~a)$ which are the parameters corresponding to the mass and spin of the central black hole. The parameter estimation errors are
\begin{align}
\Delta \lambda_i\approx\sqrt{(\Gamma^{-1})_{ii}}
\end{align}

Assuming that the spin is 0.5 and the masses of the SMBH (Sgr A*) and the small body are $4.3\times 10^6 ~ {\rm M_\odot}$ and 0.1 ${\rm M_\odot}$, by generating the plunging waveforms from the Teukolsky equation, we find that the mass and spin of the central black hole can be measured to within a fractional error of
\begin{align}
	\Delta M/ M\approx 2 \times 10^{-5} \,, \\
    \Delta a/a \approx 1 \times 10^{-4} \,.
\end{align}
This high accuracy is due to the very high SNR for the Galactic sources. For source located at a few million parsecs, the plunging burst signal can still constrain the spin of a central black hole to an accuracy of about a few per cent.

%\red{We can see it will determine the parameters of central black hole in a high accuracy.}

%\section{§3. Event rates of XMRBs in our Galaxy}

\section{ Event rates of XMRBs in our Galaxy}

We can estimate the event of these bursts using the formula \citep{xmri1},
\begin{align}
	\Gamma\simeq \frac{N}{T_{\rm rlx}
	\ln(\theta_{\rm lc}^{-2})}\label{rate}
\end{align}
where $N$ is the number of brown dwarfs, $T_{\rm rlx}$ is the relaxation timescale due to star-star scattering (two-body relaxation), and $\theta^2_{\rm lc}$ is the solid angle of the loss cone within which a brown dwarf will plunge into the SMBH.  Because brown dwarfs normally come from elongated orbits, we can calculate $\theta_{\rm lc}$ with $L/L_c$, where $L=8$ is the maximum angular momentum in our simulation that still leads to a significant SNR, and $L_c$ is the angular momentum of a circular orbit with the same energy as the plunging brown dwarf.

We note that the exact value of $\Gamma$ depends on the distance from the SMBH.
However, in the limit $\theta_{\rm lc}\ll 1$, the majority of the stars
in the loss cone come from within the influence radius, where the enclosed stellar mass becomes comparable to the mass of the SMBH \citep{LC2013}.  Therefore,
we can estimate $\Gamma$ using the values at the influence radius of the SMBH at the Galactic Centre, about 3 pc, and the corresponding relaxation time-scale is $T_{\rm rlx}\sim10^9$ years \citep{G2010}. To
derive $N_*$, we follow the assumption in Amaro-Soeane(2019) about the initial
mass function of stars, and find that there are about $N\sim3\times10^{6}$
brown dwarfs within the influence radius. With these consideration, we find
that $\Gamma\sim10^{-3}\,{\rm yr^{-1}}$.

Equation~(\ref{rate}) is derived under the assumption that the nuclear star
cluster around the SMBH is spherical. However, observations have shown that the stellar distribution in the central 8 pc of the Galactic Centre is triaxial \citep{FK2017}. In such a potential, the loss cone is refilled
mainly by stars on chaotic orbits, and the loss-cone filling rate can be orders of magnitude higher than the rate resulting from two-body relaxation \citep{MP2004}. For this reason, we think it possible that the event rate of the XMRBs
in the Galactic Centre could reach $\Gamma\sim10^{-2}\,{\rm yr^{-1}}$.  A
careful modeling of the XMRB rate in a triaxial potential is needed to better
quantify this.

However, the event rate estimation above is for our Galaxy. If we consider a 10 Mpc distance, assuming that the number density of SMBHs is 0.1/Mpc$^3$ for $10^5 {\rm M_\odot}$ and 0.01-0.1/Mpc$^3$ for $10^6 {\rm M_\odot}$ \citep{Marconi2004}, the number of SMBHs inside 10 Mpc is about 400-800. For this distance, from the Table \ref{tablesnr}, the SNR can achieve at $\sim 10$ for the dominant (2,2) modes. Therefore, if we take into account the neighbor galaxies inside 10 Mpc, the event rate can arrive at 4-8${\rm yr^{-1}}$, which is high enough for detection by space-borne detectors such as LISA. The detection of these types of plunging signals will reveal the nature of black holes.

%\section{§4. Conclusions}

\section{ Conclusions}

The very extreme mass-ratio sources in our Galaxy are crucial for the space detection of GWs. Owing the extremely small mass ratio, $\sim 10^{-8}$, the gravitational self-force of the small body can be ignored. This, together with the high SNR, means that the space–time of the central SMBH can be figured out very precisely.

In the present work, we have considered GW sources known as XMRBs in our Galaxy and neighboring galaxies, which have detectable event rates. Rather than considering Galactic inspiral sources \citep{xmri1,xmri2},we considered
compact objects such as brown dwarfs and primordial black holes
plunging into or being scattered by the central supermassive black
hole in Sgr A* or nearby SMBHs. Because both the plunging and
scattering of small objects into the black hole produce transient GW
signals (compared with inspiraling sources), we refer to them as very
extreme mass-ratio bursts (XMRBs). The frequency of this source is around $10^{-3}\sim 10^{-2}$ Hz, which corresponds to the most sensitive
frequency band of the space-borne detectors. This kind of signal
usually continues from a few tens of minutes to a few hours for
SMBHs with a mass $\sim 10^6 {\rm M_\odot}$.

The Galactic inspiraling objects stay outside the innermost stable circular orbit of the SMBHs, but the plunging objects collide with the black hole horizon directly, and then the latter can produce GW signals carrying direct information about the horizon. Our calculations show that the SNR is about $10^2\sim 10^3$ for the plunging XMRBs, and $\sim$ 20 for the scattering ones with angular momentum up to $L = 8$ for the sources in our Galaxy. For the source at 10 Mpc, the SNR can still be as large as $\sim 10$. The signals are strong enough to be detected and the event rate we estimated can reach 0.01 ${\rm yr^{-1}}$ for
the plunging sources in the Galaxy (the event rate is not sensitive to
angular momentum). If we are lucky, this kind of source will be very
important for observing the nearest SMBH. Our estimation shows
that LISA may be able to determine the mass and spin of the nearest
SMBH with an accuracy of better than $10^{-4}$ by observing one XMRB. However, in a radius 10 Mpc volume, there are a few hundreds galaxies \citep{Marconi2004}, and thus the event rate may reach $\sim 4 - 8$ ${\rm yr^{-1}}$, and the SNR of plunge sources is still large enough for space-borne detectors. This makes the XMRBs as potential sources for future space-borne detectors and the detection of the nature of
black holes.
%\nocite{*}

\section*{Acknowledgements}

This work is supported by NSFC No.11273045, and we appreciate the anonymous Referee’s suggestions about our work. This work was also supported by MEXT, JSPS Leading-edge Research Infrastructure Program, JSPS Grant-in-Aid for Specially Promoted Research 26000005, JSPS Grant-in-Aid for Scientific Research on Innovative Areas 2905: JP17H06358, JP17H06361 and JP17H06364, JSPS Core-to-Core Program A. Advanced Research Networks, JSPS Grant-in-Aid for Scientific Research (S) 17H06133, the joint research program of the Institute for Cosmic Ray Research, University of Tokyo, and by Key Research Program of Frontier Sciences, CAS, No. QYZDB-SSW-SYS016.

\begin{thebibliography}{mnras}
\bibitem[\protect\citeauthoryear{Abbott et al.}{2016a}]{gw15a} Abbott B.~P., et al., 2016, PhRvL, 116, 061102
\bibitem[\protect\citeauthoryear{Abbott et al.}{2016b}]{gw15b}Abbott B.~P., et al., 2016, PhRvL, 116, 241103

\bibitem[\protect\citeauthoryear{Abbott et al.}{2017b}]{gw17b} Abbott B.~P., et al., 2017, PhRvL, 119, 141101

\bibitem[\protect\citeauthoryear{Abbott et al.}{2017a}]{gw17a}Abbott B.~P., et al., 2017, PhRvL, 118, 221101

\bibitem[\protect\citeauthoryear{Amaro Seoane et al.}{2007}]{second_editor_1}
Amaro-Seoane P., Gair J.~R., Freitag M., Miller M.~C., Mandel I., Cutler C.~J., Babak S., 2007, CQGra, 24, R113

\bibitem[\protect\citeauthoryear{Amaro Seoane}{2018}]{emris} Amaro-Seoane P., 2018, LRR, 21, 4

\bibitem[\protect\citeauthoryear{Amaro Seoane}{2019}]{xmri1}  Amaro-Seoane P., 2019, PhRvD, 99, 123025


\bibitem[\protect\citeauthoryear{Arfken}{1985}]{Green} Arfken G., Mathematical Methods for Physicists (Academic Press, Orlando, 1985), chapter 16.

\bibitem[\protect\citeauthoryear{Babak et al.}{2017}]{second_editor_2}
Babak S., et al., 2017, PhRvD, 95, 103012

\bibitem[\protect\citeauthoryear{Berry \& Gair}{2013a}]{berry13a} Berry C.~P.~L., Gair J.~R., 2013, ASPC, 467, 185, ASPC..467

\bibitem[\protect\citeauthoryear{Berry \& Gair}{2013b}]{berry13b} Berry C.~P.~L., Gair J.~R.,
  %``Extreme-mass-ratio-bursts from extragalactic sources,''
 2013, MNRAS, 433, 3572

\bibitem[\protect\citeauthoryear{Berry et al.}{2019}]{second_editor_3}
Berry C., et al., 2019, BAAS, 51, 42

\bibitem[\protect\citeauthoryear{Cutler \& Flanagan}{1994}]{Cutler1994} Cutler C., Flanagan {\'E}. E., 1994, PhRvD, 49, 2658

\bibitem[\protect\citeauthoryear{Danzmann et al.}{2017}]{lisal3} Danzmann K., et. al., 2017,  A proposal in response to the ESA call for L3 mission
concepts

\bibitem[\protect\citeauthoryear{Feldmeier Krause et al.}{2017}]{FK2017} Feldmeier-Krause A., Zhu L.,Neumayer  N., et al.,
2017, MNRAS, 466, 4040


\bibitem[\protect\citeauthoryear{Fern{\'a}ndez \& Kobayashi}{2019}]{2019MNRAS} Fern{\'a}ndez J.~J., Kobayashi S., 2019, MNRAS, 487, 1200


\bibitem[\protect\citeauthoryear{Fujita \& Tagoshi}{2004}]{MST_r}
 Fujita R., Tagoshi H., 2004, PThPh, 112, 415

\bibitem[\protect\citeauthoryear{Fujita \& Tagoshi}{2005}]{MST_c}
Fujita R., Tagoshi H., 2005, PThPh, 113, 1165

\bibitem[\protect\citeauthoryear{Genzel et al.}{2010}]{G2010}Genzel R., Eisenhauer F., Gillessen S., 2010, RvMP, 82, 3121

\bibitem[\protect\citeauthoryear{Gourgoulhon et al.}{2019}]{xmri2}Gourgoulhon E., Le Tiec A., Vincent F.~H., Warburton N., 2019, A\&A, 627, A92

\bibitem[\protect\citeauthoryear{Han}{2009}]{PN} Han W.-B., 2009, IJTP, 48, 621

\bibitem[\protect\citeauthoryear{Han}{2010}]{han10} Han W.-B., 2010, PhRvD, 82, 084013

\bibitem[\protect\citeauthoryear{Han et al.}{2017}]{han17} Han W.-B., Cao Z., Hu Y.-M., 2017, CQGra, 34, 225010


\bibitem[\protect\citeauthoryear{Hopman et al.}{2007}]{hopman07} Hopman C., Freitag M., Larson S.~L., 2007, MNRAS, 378, 129

\bibitem[\protect\citeauthoryear{http://lisa.nasa.gov/}{}]{lisa}
http://lisa.nasa.gov/

\bibitem[\protect\citeauthoryear{Hughes}{2000}]{Hughes 2000} Hughes S.~A., 2000, PhRvD, 62, 044029

\bibitem[\protect\citeauthoryear{Hughes}{2000}]{t-s1} Hughes S.~A., 2002, PhRvD, 65, 069902

\bibitem[\protect\citeauthoryear{Hu \& Wu}{2017}]{hu2017the}
 Hu W.-R. and Wu Y.-L., 2017,
  %``The Taiji Program in Space for gravitational wave physics and the nature of gravity,''
  Natl.\ Sci.\ Rev.\  {\bf 4}, no. 5, 685
  %doi:10.1093/nsr/nwx116
  %%CITATION = doi:10.1093/nsr/nwx116;%%
  %38 citations counted in INSPIRE as of 30 Mar 2020

\bibitem[\protect\citeauthoryear{Leaver}{1985}]{Leaver 1985}  Leaver E.~W., 1985, RSPSA, 402, 285

\bibitem[\protect\citeauthoryear{Linial \& Sari}{2017}]{2017MNRAS} Linial I., Sari R., 2017, MNRAS, 469, 2441

\bibitem[\protect\citeauthoryear{Liu \& Chen}{2013}]{LC2013}Liu F.~K., Chen X., 2013, ApJ, 767, 18

\bibitem[\protect\citeauthoryear{Luo et al.}{2017}]{luo2016tianqin}
 Luo J., et al., 2016, CQGra, 33, 035010
 
\bibitem[\protect\citeauthoryear{Mano et al.}{1996}]{Teukolsky 1996}Mano S., Suzuki H., Takasugi E., 1996, PThPh, 95, 1079

\bibitem[\protect\citeauthoryear{Marconi et al.}{2004}]{Marconi2004} Marconi A., Risaliti G., Gilli R., L. K. Hunt, Maiolino R., Salvati M.,
%``Local Supermassive Black Holes, Relics of Active Galactic Nuclei and the X-ray Background,''
2004, MNRAS, 351, 169

\bibitem[\protect\citeauthoryear{Merritt \& Poon}{2004}]{MP2004}Merritt D., Poon M.~Y., 2004, ApJ, 606, 788

\bibitem[\protect\citeauthoryear{Mino et al.}{1997}]{t-s2} Mino Y., Sasaki M., Shibata M., Tagoshi H., Tanaka T., 1997, PThPS, 128, 1

\bibitem[\protect\citeauthoryear{Moore et al.}{2015}]{SNR}Moore C.~J., Cole R.~H., Berry C.~P.~L., 2015, CQGra, 32, 015014


\bibitem[\protect\citeauthoryear{Peters \& Mathews}{1963}]{peters 1963} Peters P.~C., Mathews J., 1963, PhRv, 131, 435

\bibitem[\protect\citeauthoryear{Peters}{1964}]{peters 1964}Peters P.~C., 1964, PhRv, 136, 1224

\bibitem[\protect\citeauthoryear{Regge \& Wheeler}{1957}]{RW1957} Regge T., Wheeler J.~A., 1957, PhRv, 108, 1063

\bibitem[\protect\citeauthoryear{Rubbo et al.}{2006}]{rubbo06}Rubbo L.~J., Holley-Bockelmann K., Finn L.~S., 2006, AIPC, 873, 284, AIPC..873

\bibitem[\protect\citeauthoryear{Ruffini \& Sasaki}{1981}]{1981Ruffini} Ruffini R., Sasaki M., 1981, PThPh, 66, 1627

\bibitem[\protect\citeauthoryear{Sasaki \& Nakamura}{1982}]{SN}Sasaki M., Nakamura T., 1982, PThPh, 67, 1788

\bibitem[\protect\citeauthoryear{Teukolsky}{1973}]{Teukolsky}Teukolsky S.~A., 1973, ApJ, 185, 635


\bibitem[\protect\citeauthoryear{Toonen et al.}{2009}]{toonen09} Toonen S., Hopman C., Freitag M., 2009, MNRAS, 398, 1228

\bibitem[\protect\citeauthoryear{Yunes, et al.}{2008}]{2008ApJ} Yunes N., Sopuerta C.~F., Rubbo L.~J., Holley-Bockelmann K., 2008, ApJ, 675, 604

\bibitem[\protect\citeauthoryear{Zerilli}{1970}]{Z1970} Zerilli F.~J., 1970, PhRvD, 2, 2141

\end{thebibliography}

%%%%%%%%%%%%%%%%%%%%%%%%%%%%%%%%%%%%%%%%%%%%%%%%%%

%%%%%%%%%%%%%%%%% APPENDICES %%%%%%%%%%%%%%%%%%%%%

%\appendix



%%%%%%%%%%%%%%%%%%%%%%%%%%%%%%%%%%%%%%%%%%%%
\bsp	% typesetting comment
\label{lastpage}

\end{document}